\documentclass[%
 reprint,
superscriptaddress,
showpacs,
preprintnumbers,
 amsmath,amssymb,
 pre,
]{revtex4-1}
\usepackage{setspace}
\usepackage{lipsum}
\usepackage{graphicx}
\usepackage{wasysym} 
\usepackage{dcolumn}
\usepackage{bm}
\usepackage{caption}

\usepackage{color}

\usepackage[english]{babel}
\usepackage[utf8]{inputenc}

\begin{document}

\title{
Information parity in complex networks
}

\author{Aline Viol}
 \email{aline.viol@bccn-berlin.de}
 \affiliation {Institute of Theoretical Physics, Technische Universit\"at Berlin, Hardenbergstra\ss{}e 36, 10623 Berlin, Germany}
\affiliation{Bernstein Center for Computational Neuroscience Berlin, Humboldt-Universit{\"a}t zu Berlin, Philippstra{\ss}e 13, 10115 Berlin, Germany}

\author{Vesna Vuksanovi{\'c}}
\affiliation {Aberdeen Biomedical Imaging Centre, University of Aberdeen, Foresterhill, Aberdeen AB25 2ZD, UK}

 \author{Philipp H\"ovel}
 \email{philipp.hoevel@ucc.ie}
  \affiliation {Institute of Theoretical Physics, Technische Universit\"at Berlin, Hardenbergstra\ss{}e 36, 10623 Berlin, Germany}
   \affiliation{Bernstein Center for Computational Neuroscience Berlin, Humboldt-Universit{\"a}t zu Berlin, Philippstra{\ss}e 13, 10115 Berlin, Germany}
  \affiliation {School of Mathematical Sciences, University College Cork, Western Road T12 XF62, Cork, Ireland}

\date{\today}

\begin{abstract}

\noindent A growing interest in complex networks theory results in an ongoing demand for new analytical
tools. We propose a novel measure based on information theory that provides a new perspective
for a better understanding of networked systems: Termed ``information parity,” it quantifies the
consonance of influence among nodes with respect to the whole network architecture. Considering the statistics of geodesic distances, information parity detects how similarly a pair of nodes can influence and be influenced by the network.This allows us to quantify the quality of information gathered by the nodes. To demonstrate the method’s potential, we evaluate a social network and human brain networks. Our results indicate that emerging phenomena like an ideological orientation of nodes in social network is severely influenced by their information parities. We also show that anatomical brain networks have a greater information parity in inter-hemispheric homologous regions placed near the midsagittal plane. Finally, functional networks have, on average, greater information parity for inter-hemispheric homologous regions in comparison to the whole network. We find that a pair of regions with high information parity exhibits higher correlation, suggesting that the functional correlations between cortical regions can be partially explained by the symmetry of their overall influences of the whole brain.

\end{abstract}


\maketitle

%
\section{Introduction}

Within the last few decades, complex network approaches have pervaded a number of scientific fields. 
This interest expanded, in part, by virtue of technological advances
that acquire novel datasets such as brain imaging techniques and online
communication networks \cite{RUB10a,Vespignani425}.
Besides much progress towards a general understanding of complex systems, the popularity of network modeling has resulted in a growing demand for new analytical methods. 
We use notions of the information theory \cite{shannon48} to create a new method for analyzing the influence of the network topology on nodes' behavior. 
The study of networks using concepts of information theory has increased in literature \cite{VIO17a,Anand09,Tost09,honey08}. 
	The main assets of the method proposed in this paper are (i) the simplicity - it is based on an established metric of the network topology; (ii) the clear and interpretable meaning of the output and the convenience for experimental studies; (iii) the novelty in regards to an understanding of the overall influence of network topology on its nodes. 

We are driven by the following questions: 
How are the nodes of a network influenced or influencing the overall structure? Does the overall network topology affect the nodes' role and local behaviors? %
We tackle these questions quantifying the similarity of influence patterns between pairs of nodes.  
In a networked system, the elements influence each other directly via links connecting them and indirectly via nodes that are two or more links away.
The diversity of influences can be assessed by the geodesic distance matrix. In that matrix, the rows record the relative position of a node to the others \cite{NEW10}.
Hence, the geodesic distances matrix is a key object to evaluate the symmetry of influences in networks  \cite{ge}. 
Inspired by mutual information \cite{shannon48}, the measure proposed in this paper, termed ``information parity,"  uses the distribution of geodesic distances to quantify the similarity of influence of pairs of nodes. In other words, information parity quantifies to what extent one can infer the influences of a node given the knowledge about influences of another.
We use the general term influences because the meaning of relations depends on the nature of the links.
For example, if the links represent a channel of communication, the information parity reflects how similar is the information received for two nodes in the network and how similar is their impact on the whole network. 

We define the information parity formula in section~\ref{methods}. In section~\ref{sec:results}, we illustrate the potential of this measure by evaluating empirical networks. 
There have been many efforts to understand the phenomena of social ideological polarization \cite{INTERIAN18,DelVicario2017,Bail18}, that is, when a community splits in two groups with different opinions about one or more issues.
Our results suggest that individual opinion in a polarized network are strongly influenced by the whole networks topology.
Subjects -- represented as nodes -- with high information parity tend to choose the same side of polarization.
Besides, we demonstrate that information parity can also bring valuable insight into studies of human brain networks unveiling patterns on anatomical 
and functional brain networks. Decoding network patters of brain networks is one of the major challenges of contemporary neuroscience.
\vspace{1cm}

\section{Methods}
\label{methods}
Let us define a unweighted and undirected network $G(V,E)$, where $V$ is a set of $N$ nodes and $E$ is the set of their links.
The adjacency matrix $A$ is defined here as $A_{ij}=1$ if a pair of nodes are connected and $A_{ij}=0$ otherwise.
Links describe relations of different natures depending on the problem. Since it is often impossible to find a spatial embedding of networks and corresponding Cartesian coordinates, the geodesic distance is defined by the shortest number of links along a path leading from a specific node to another \cite{NEW10}.
The geodesic distance matrix reflects the network topology. Each row of the geodesic distance matrix comprises the relationship of a node with all nodes of the network; it shows how the whole network is seen from that node's perspective \cite{ge}.
%
%
We define $p_i(r)$ as the probability of finding a node in the network at a distance $r$ from the node $i$ and $p_{ij}(r)$ as the probability of finding a node at a distance $r$ from a pair of nodes $i$ and $j$: 
\begin{equation}
  p_{i} (r)= \frac{1}{N-1}\sum_{\substack{ k \in{V} \\ k\neq i}}{ \delta_{D_{ik},r} } 
\end{equation}
\begin{equation}
  p_{ij}(r)= \frac{1}{N-2}\sum_{\substack{ k \in V \\ k\neq i,j}} \delta_{D_{ik},r} \, \delta_{D_{jk},r}  ~,
\end{equation}
where $\delta_{\cdot,\cdot}$ denotes the Kronecker delta, $\{D_{ij}\}_{i,j=1,\cdots,N}$ is the geodesic distance matrix,
and the parameter $r$ assumes integer values in the interval $1\leq r\leq r_{\textrm{max}}$ with $r_{\textrm{max}}$ being the nodes' maximum neighborhood radius \cite{ge}.  
The information parity is defined as:
\begin{equation}  
\label{formula}
  I_{ij}=  \sum_{r=1}^{r_{\textrm{max}}} p_{ij}(r)  \log\frac{{p_{ij}(r)}}{p_i(r)p_j(r)}.
\end{equation}
This formula differs from mutual information \cite{shannon48} because it does not consider a joint probably of two random variables. Instead, it considers the probability of two nodes being equidistant to the same node. Unlike mutual information, the information parity may assume negative values. In that case, the pair of nodes evaluated does not have a considerable amount of common influences corresponding to a high geodesic entropy \cite{ge}. The negative values can be considered as a ``information disparity."     

Information parity quantifies the similarity of influences for a pair of nodes taking into account their relative positions in the network structure. One can also consider $r_i$ and $r_j$ in the definition of the probability $p_{ij}(r_i,r_j)$ as two independent parameters. Then, $I_{ij}$ quantifies how the influences of $i$ is constrained to the influences $j$, and vice versa. See supplementary material for further information. 
In this paper, we focus on the special case where the distances are constrained $r_i=r_j=r$, that is, we consider the network neighborhoods equidistant to nodes $i$ and $j$.
This simplified case has an intuitive meaning: the higher the information parity of a pair of nodes, the more similar are their interaction patterns in the network.  

In order to illustrate the capability of the influence symmetry, we address two relevant problems in networks science:
ideological polarization in social networks and bilateral symmetry in human brain networks. The approach can be easily transferred for other contexts as well as generic classes of networks.

\section{Results}
\label{sec:results}
%
\subsection{Social network with ideological polarization}
We evaluate the information parity of a small and well-documented social network known as Karate-club network, firstly studied by Zachary \cite{zachary}. This network have been extensively explored in the literature, becoming a classical example of an ideologically polarized social network. In short, a disagreement between the president of the club and one of the instructors led to an ideological polarization resulting later in a rupture of the club. One part of the members left along with the instructor, Mr. Hi, and the other part stayed in the original club \cite{zachary}. The topology of the Karate-club network was defined by Zachary according to the personal relationships between the karate-club's members. He proposed a model to detect the polarization considering the structure including the strength of connections between members \cite{zachary}. Later, Girvan and Newman demonstrated that the polarization could be detected considering only the structure of connections \cite{Girvan02}.
Using the unweighted and undirected network, we show that the individual decision, about which group to join, is reflected by the information parity.
\begin{figure*}[ht]
    (a)    
	\includegraphics[width=0.45\linewidth]{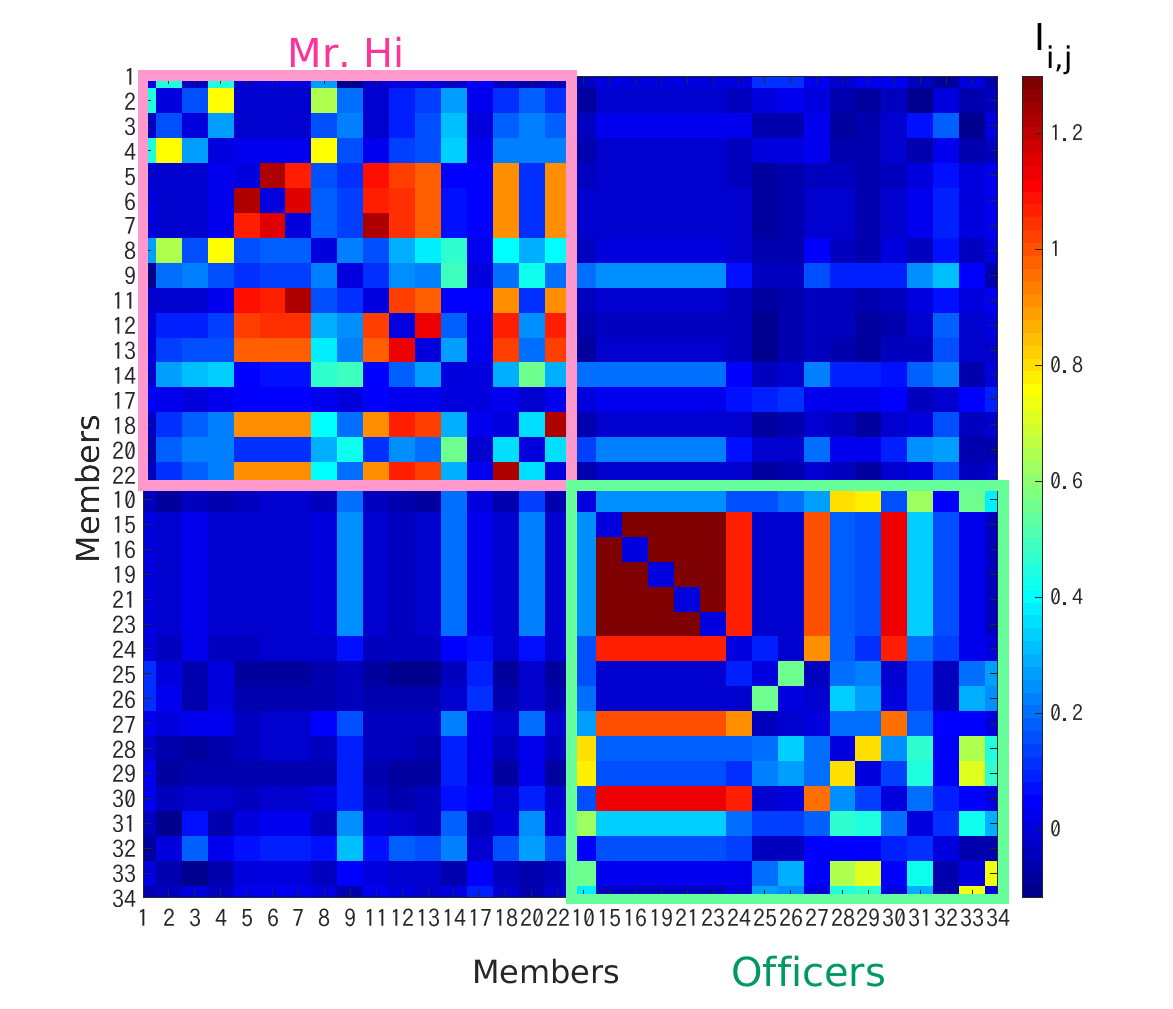}
	(b)
	\includegraphics[width=0.45\linewidth]{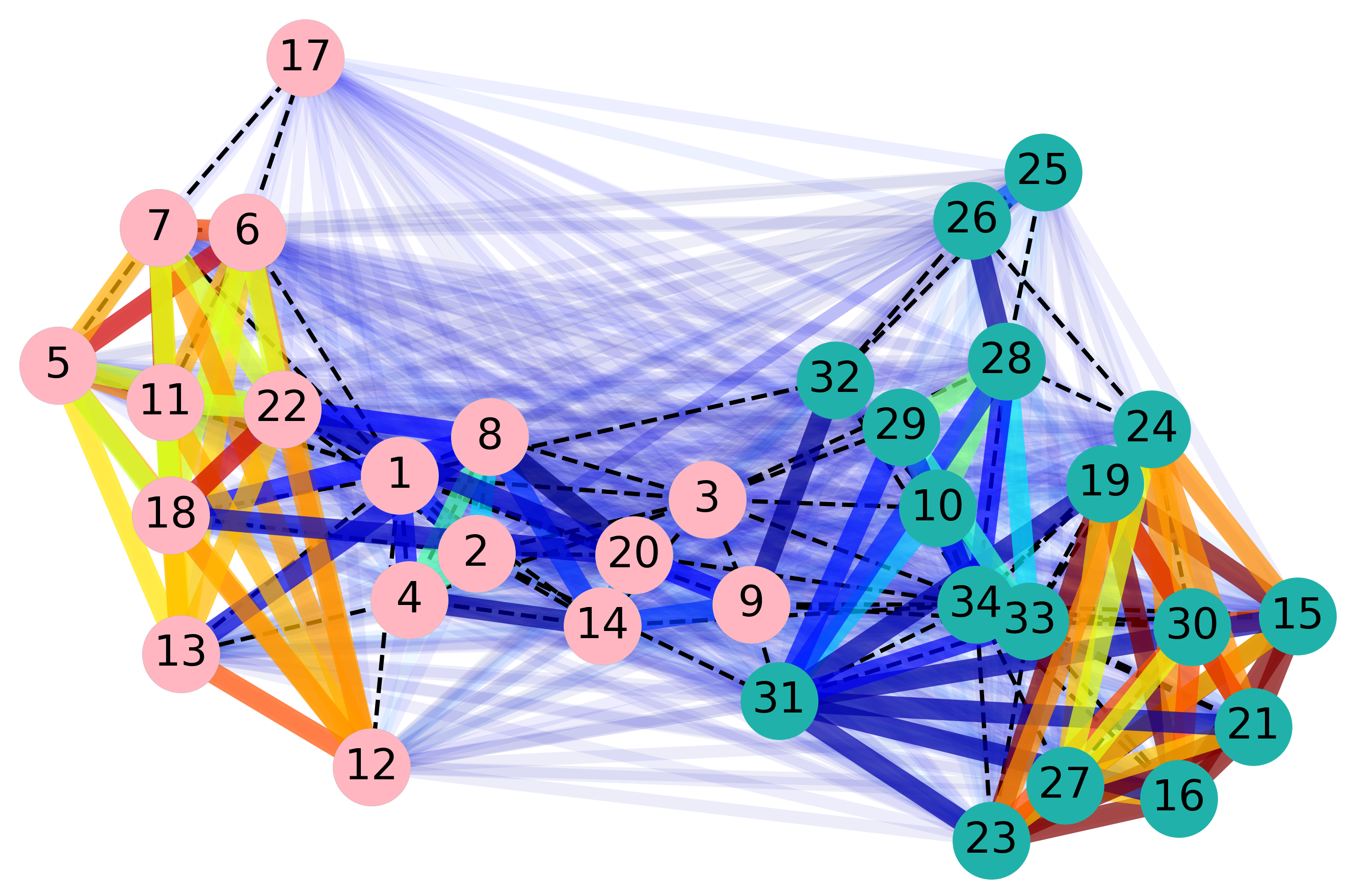}    
	\caption{Information parity of the Karate-club network. (a) Information parity matrix sorted according to the group that each of the 34 members belong after the fission keeping the labels from the original study.  
	(b) Schematic illustration showing the information parity on the network. Connections with $I_{ij}<0.3$ are shaded to improve the visualization.  Members from Mr. Hi and the Officers group are represented by pink and green nodes, respectively. The network links are represented in black-dashed line.}
	 \label{kcmi}
\end{figure*}  

Figure~\ref{kcmi} (a) shows the information parity matrix of the 
Karate-club network. The matrix elements refer to the information parity between pairs of members and are reordered according to the two groups for better visualization. Note that the information parity between members of the same group is greater than between members of the opposed group. These relations are depicted by colored connections on top of the original network (dashed links) in panel (b). 

\begin{figure*}[ht]   
	(a)                                   
	\includegraphics[width=0.8\linewidth]{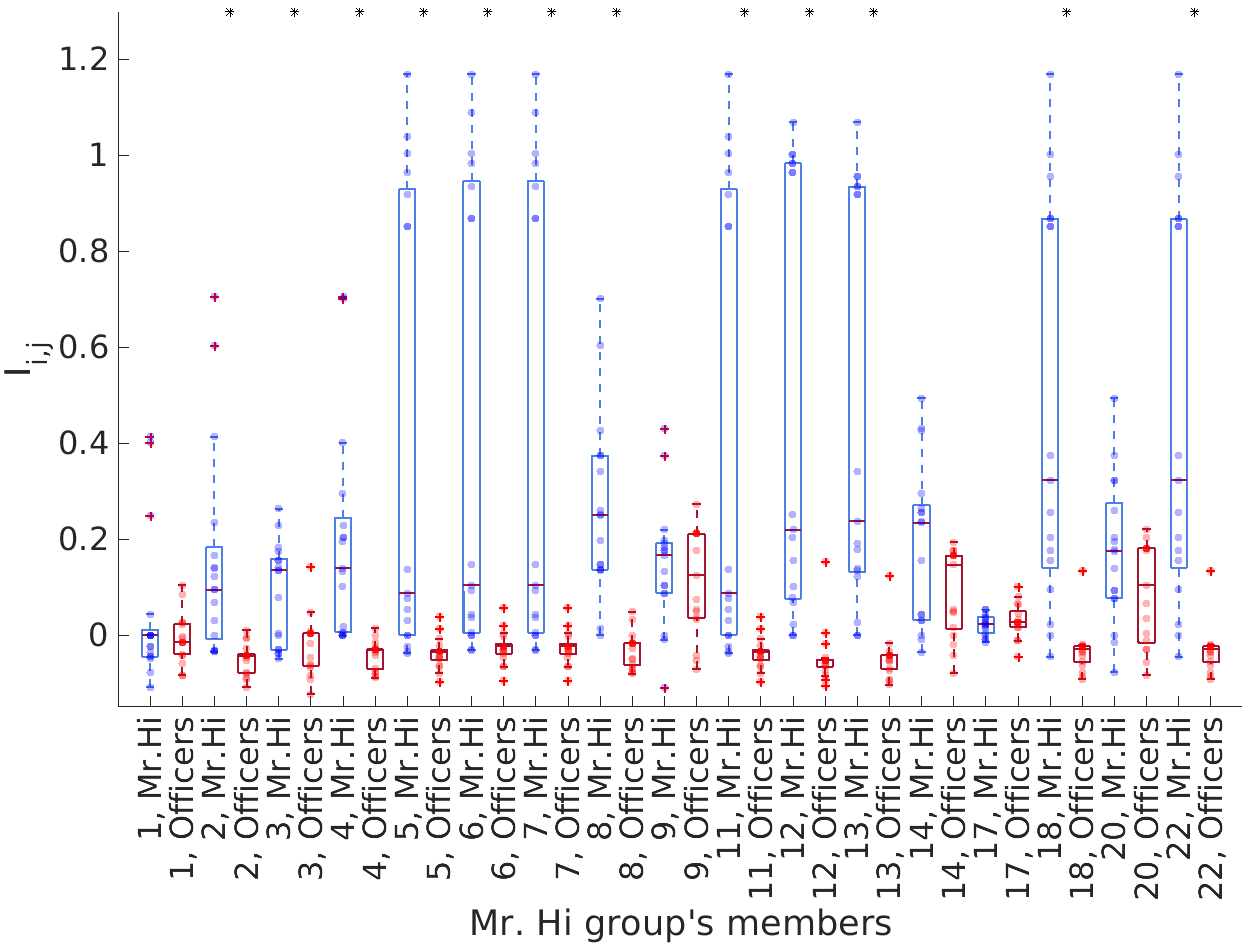}
	\vspace{0.5cm}
	
	(b)	
	\includegraphics[width=0.8\linewidth]{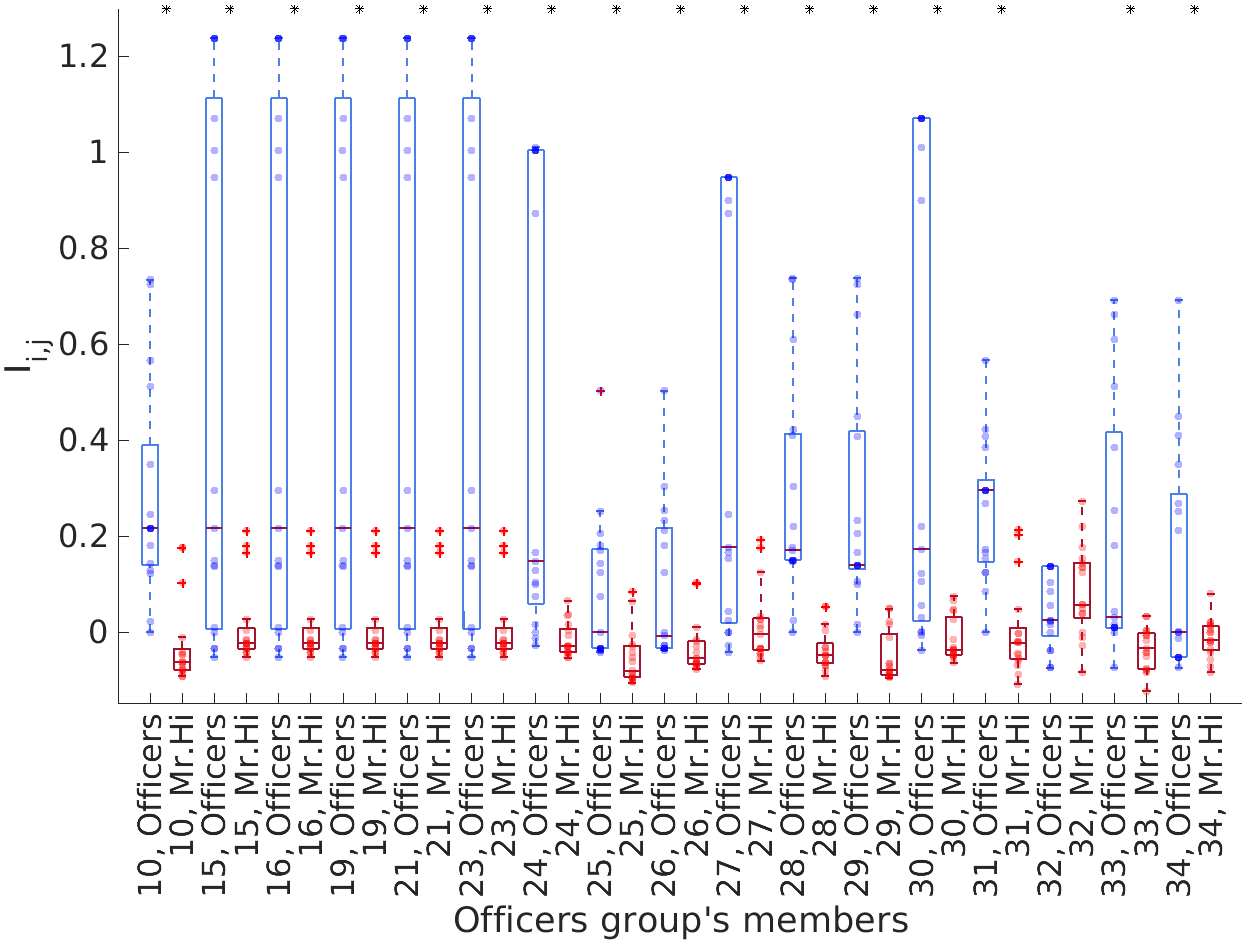}
	\caption{Comparing information parity of members inside and outside their groups. The blue boxplot depicts the information parity of each member with the members of the same group and the red boxplot with the members of the opposite group for (a) Mr. Hi group members and (b) officers group members.
		The dots inside the boxplots show the pairwise relations. The asterisk * marks the significance $p<0.05$ of the difference measure by the Student's t-test. 
	}
	\label{kcboxplot}
\end{figure*}

Figure~\ref{kcboxplot} compares the distribution of information parity within the same group (blue) and between the two groups (red) for members of Mr. Hi group and Officers group in panels (a) and (b), respectively. Note that, for the subjects where the information parity is not greater inside its group (members/nodes 17 and 32), the difference between inter- and intra-group parity is not significant. Similarly the difference for subject 9, which is a subject miss-classified by Zachary \cite{zachary}, is not significant either.
This example shows that the measure of information parity can help to understand the functioning and limitations of communities detection algorithm \cite{Newman2004}. 

The distribution of neighbors of a node in dependence on the neighborhood radius reflects the diversity of influences concerning this node \cite{ge, nico2013}. If two members share exactly the same structure of influences, that is, they are 
equidistant to the same individuals, they receive the same quality of information and tend to have similar perception and opinions. This explains why members with high information parity chose the same side after the fission, although they might not have a direct link; they are exposed to similar information which impacted their decision. 
Therefore, information parity is a valuable resource for creating strategies to analyze the influence in social groups, e.g., with respect to ideological polarization.
One can use the knowledge of information parity to control the information flow on a network, for instance, by introducing strategic connections to reduce network polarization via a few local interventions.

\subsection{Bilateral symmetry on anatomical brain networks}
\begin{figure*}[ht]
	(a)
	\includegraphics[width=0.455\linewidth]{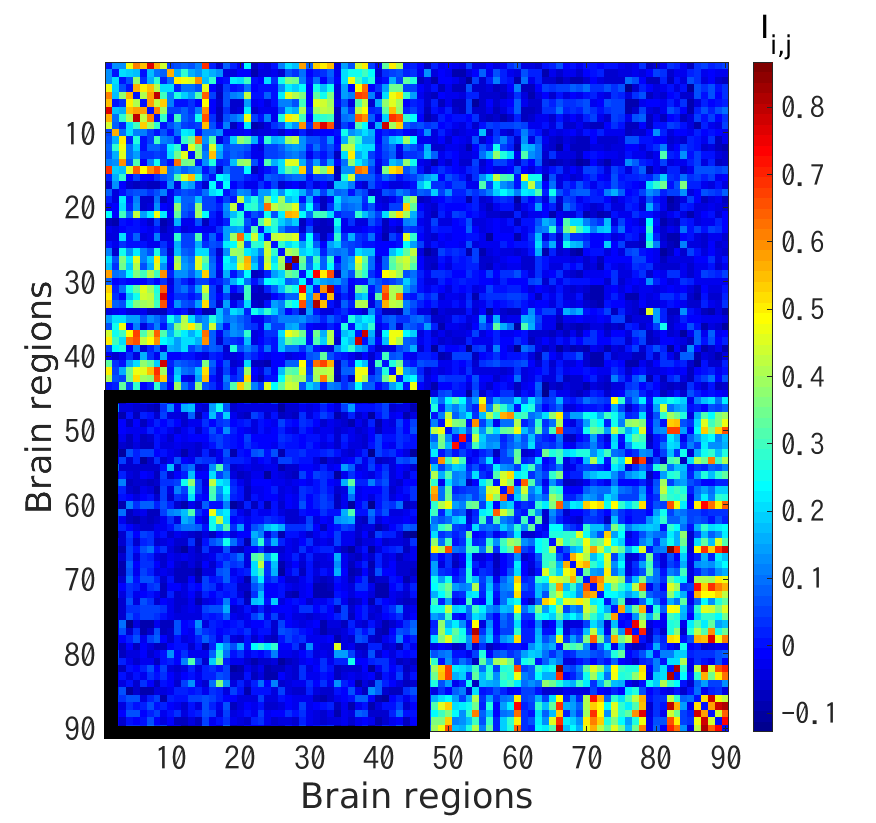}
	(b)
	\includegraphics[width=0.455\linewidth]{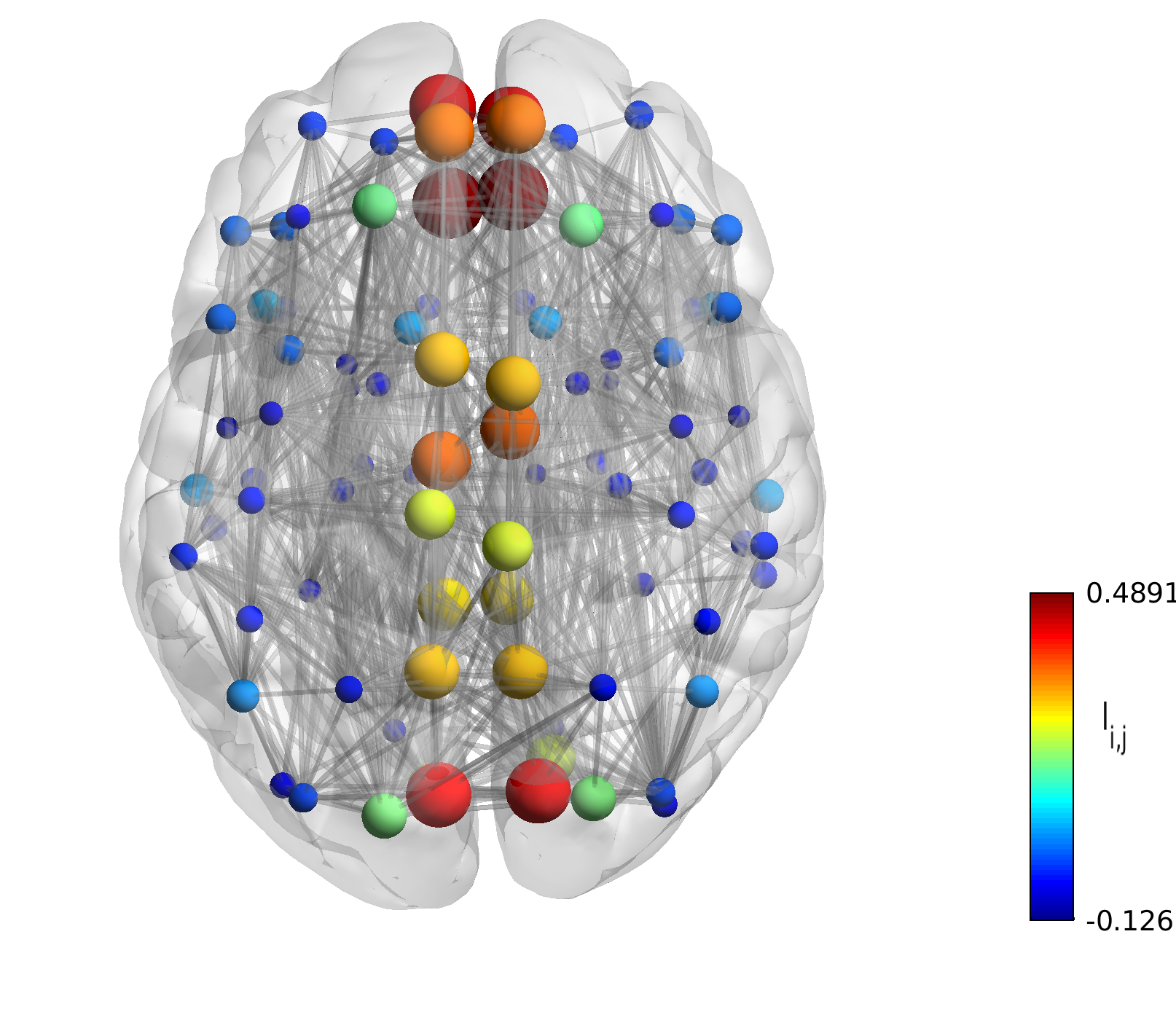} \\
	\hspace{-0.7cm}
	(c)
	\includegraphics[width=0.95\linewidth]{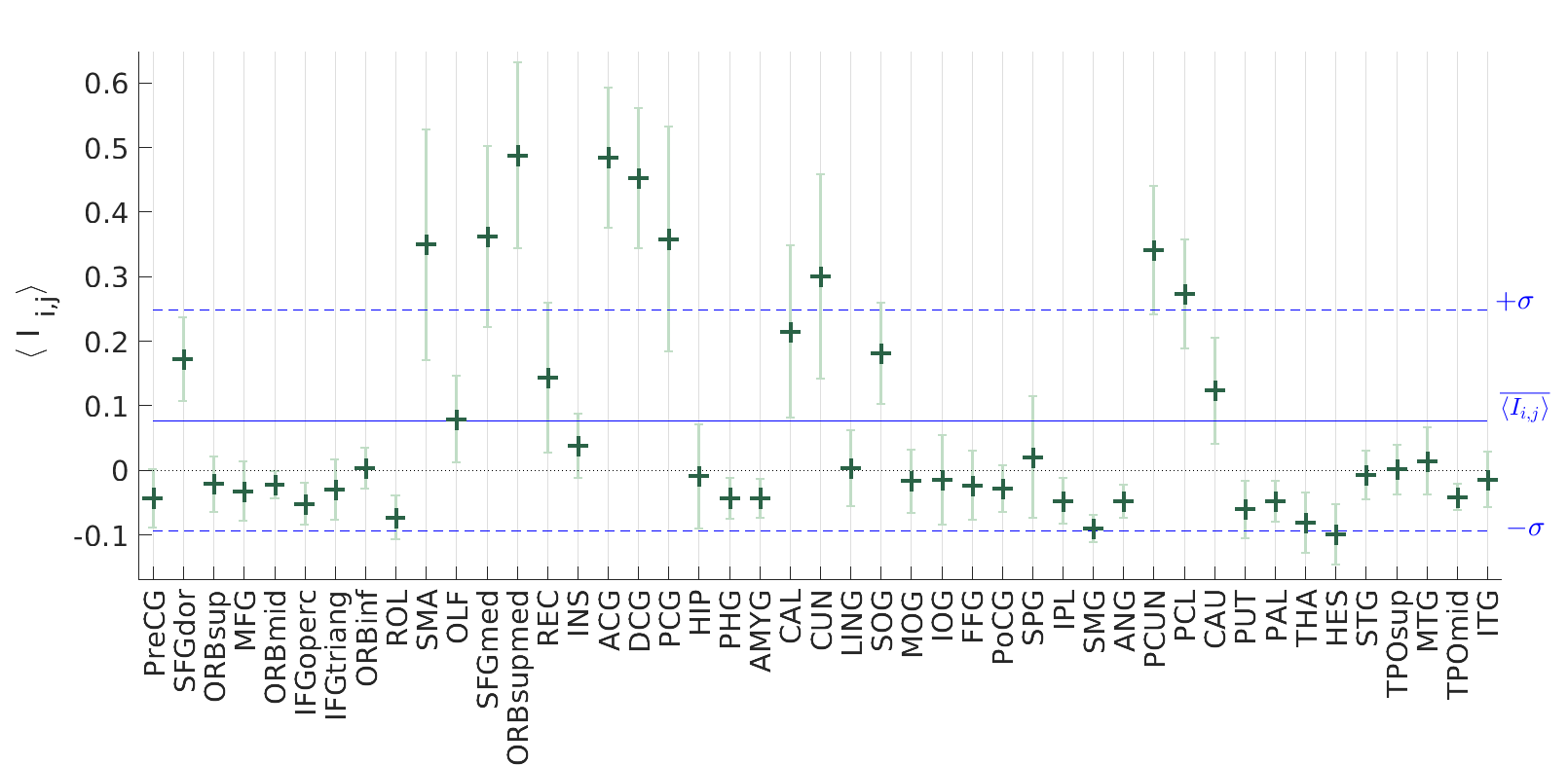}
	\caption{Information parity on anatomical brain networks. (a) Exemplary information parity matrix of one of the subjects. The mirrored inter-hemispheric regions correspond to the diagonal in the lower left quadrant (black square). (b) Information parity of the inter-hemispheric homologous regions. The size and color representing the magnitude of the information parity (visualized with the BrainNet Viewer \cite{brainnet}).  (c) Average over 21 subjects of the inter-hemispheric homologous regions with the standard deviation represented by the green bar. The blue dashed lines represent the global standard deviation $\sigma$. See supplementary material for the brain regions description.     
	}
	\label{anatomical}	
\end{figure*}

As a second example, we evaluate the information parity between human brain cortical regions with inter-hemispheric spatial correspondence regarding the structure of the anatomical connections. The data evaluated here was firstly studied by Kahn et al. \cite{Kahn16}. See supplementary material for further information.
Anatomical networks are generated from structural connectivity maps built by tracking the white matter fibers linking cortical regions \cite{Sporns2005}.
High information parity between cortical regions indicates that they potentially have similar influences taking into account the whole structure of the whole brain.
Figure \ref{anatomical} (a) shows the information parity matrix of one subject with the regions reordered according to right and left hemispheres. The illustration in panel (b) depicts the magnitude of the information parity between inter-hemispheric homologous cortical regions expressed by the color and size of the nodes. Note that the information parity of cortical regions situated on the mid-sagittal plane is significantly higher than the others. In panel (c) we compare the average information parity (over 21 subjects) of all pairs of homologous inter-hemispheric cortical regions. Several regions show negative information parity, or information disparity, indicating a strong lack of overlapping relations.
The average information parity over all possible pairs of nodes in the network is $I \approx 0.07$, considerably lower than the average for the regions on near the mid-sagittal plane. The data presented here refer to networks with mean degree $\langle k \rangle\approx 20$, but we find that the qualitative behavior does not change varying this parameter. We describe on the supplementary material the method we use to generate unweighted networks.  
The interpretation of these finding requires further analysis and is out of the scope of this article. 
One possible hypothesis is that the greater information parity near the mid-sagittal plane could be due the proximity to the corpus callosum, the largest white matter structure connecting the hemispheres \cite{HOFER2006989}.   
Another possibility is that the divergence between these regions and the other regions results from the limitation of diffusion tensor imaging (DTI) image to mapping properly all anatomical connections, in particular between the two hemispheres \cite{Sporns2005,Dougherty2005}. 



\subsection{Bilateral symmetry on functional brain networks}

Functional brain networks are derived from temporal correlation between regional activities \cite{Sporns2005}. A link in these networks denotes a statistical dependence between cortical regions signals. High correlation reflects a functional relationship according the leading paradigm on neuroscience, i.e., regions that show correlated signals support similar function \cite{HORWITZ2003466,Frith96,BAS17}.
Information parity evaluated in functional networks quantifies the overall statistical dependencies of one cortical region of the brain having information of the statistical dependencies of another. 
    
We evaluate information parity in functional human brain networks that are acquired from functional magnetic resonance images of subjects in resting state with open eyes; no frequency filter were applied. The functional matrices are the same used in the study of references \cite{Hovel2018,Vesna2016}. 
Our focus is again in the inter-hemispheric homologous cortical regions, that is, regions spatially correspondents placed on opposite hemispheres. We binarized the functional maps to define unweighted networks using a threshold method \cite{onias2014,Achard07}. See supplementary material.   

Figure~\ref{functional} (a) shows the correlation matrix of one of the subjects on left and the information parity matrix on right. The matrices are reordered to split right-left hemispheres.
Panel (b) compares the average of information parity of homologous inter-hemispheric regions with the average over all pairs of regions for each of 26 subjects. The bars represent the standard deviation considering different network densities. The average information parity between homologous regions are significantly greater when compared with the average mutual information between all cortical regions.
Panel (c) shows these two information parity values -- evaluated over the average of all subjects -- in dependence on the average degree, that is, the network density. 
The values of information parity are different for each subject due to the natural variation on functional activity among distinct subjects \cite{Finn2015}. Unlike anatomical networks, there is no significant discrepancy among the pairs regions. In particular, all evaluated pairs of bilaterally homologous regions show relative high information parity.   
 We argue that high information parity between cortical regions could 
contribute to an increase of correlation of their signals. In other words, the symmetry on the overall statistical dependencies influences the local functional connectivity.
Panel (d) relates the Pearson correlation and the information parity of 
all pairs of regions. Note that nodes with high information parity tends to have high correlation. The gray plot depicts the information parity considering only the first neighbors, that is, only the direct influences. A linear relation is also observed in this case with a smaller slope.
The influence of common first neighbors on the correlation between cortical 
regions signals has previously been suggested in synchronization models of brain 
networks \cite{VUKSANOVIC2014,Hovel2018}.
Our analysis corroborates with these results. Moreover, we extend them; the 
correlation is not only influenced by the common first neighbors but by the 
 symmetry of neighbors considering the overall structure.   

\begin{figure*}[ht]
	\centering
	(a)
	\includegraphics[width=0.41\linewidth]{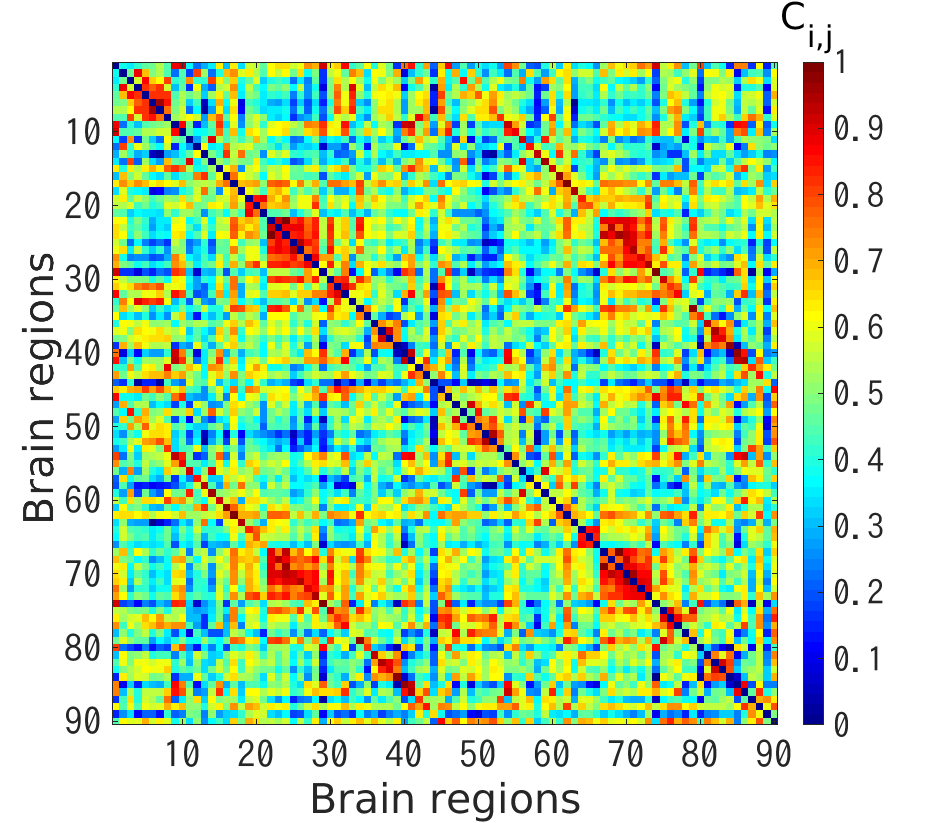}
	\includegraphics[width=0.41\linewidth]{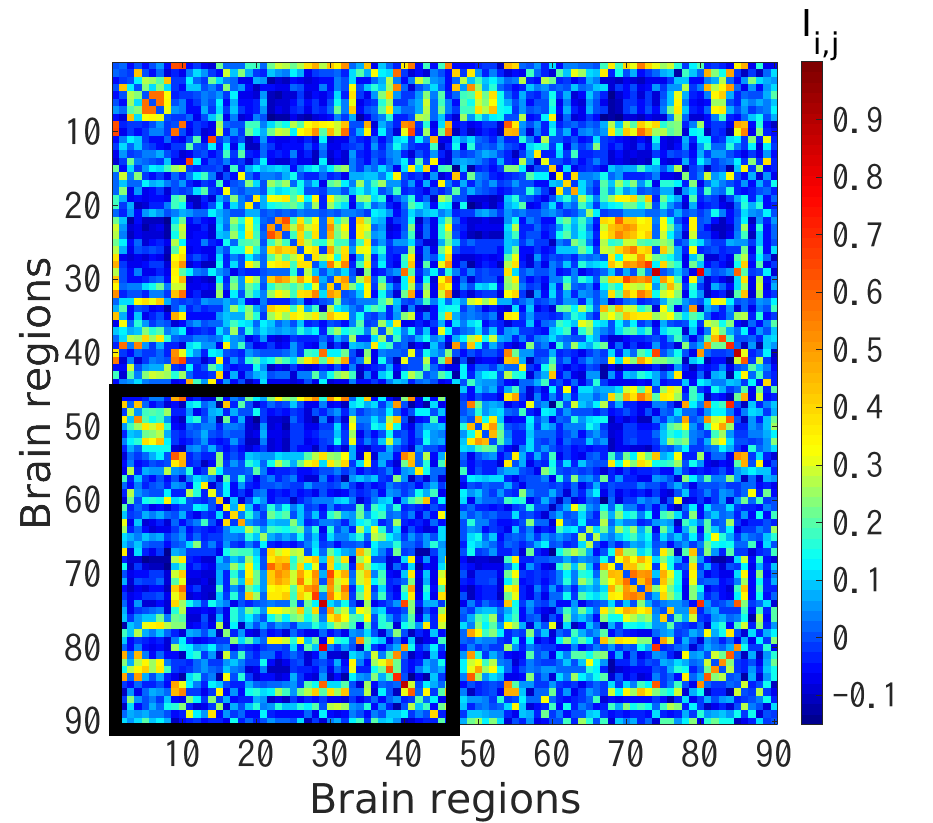}\\
	(b)
	\includegraphics[width=0.84\linewidth]{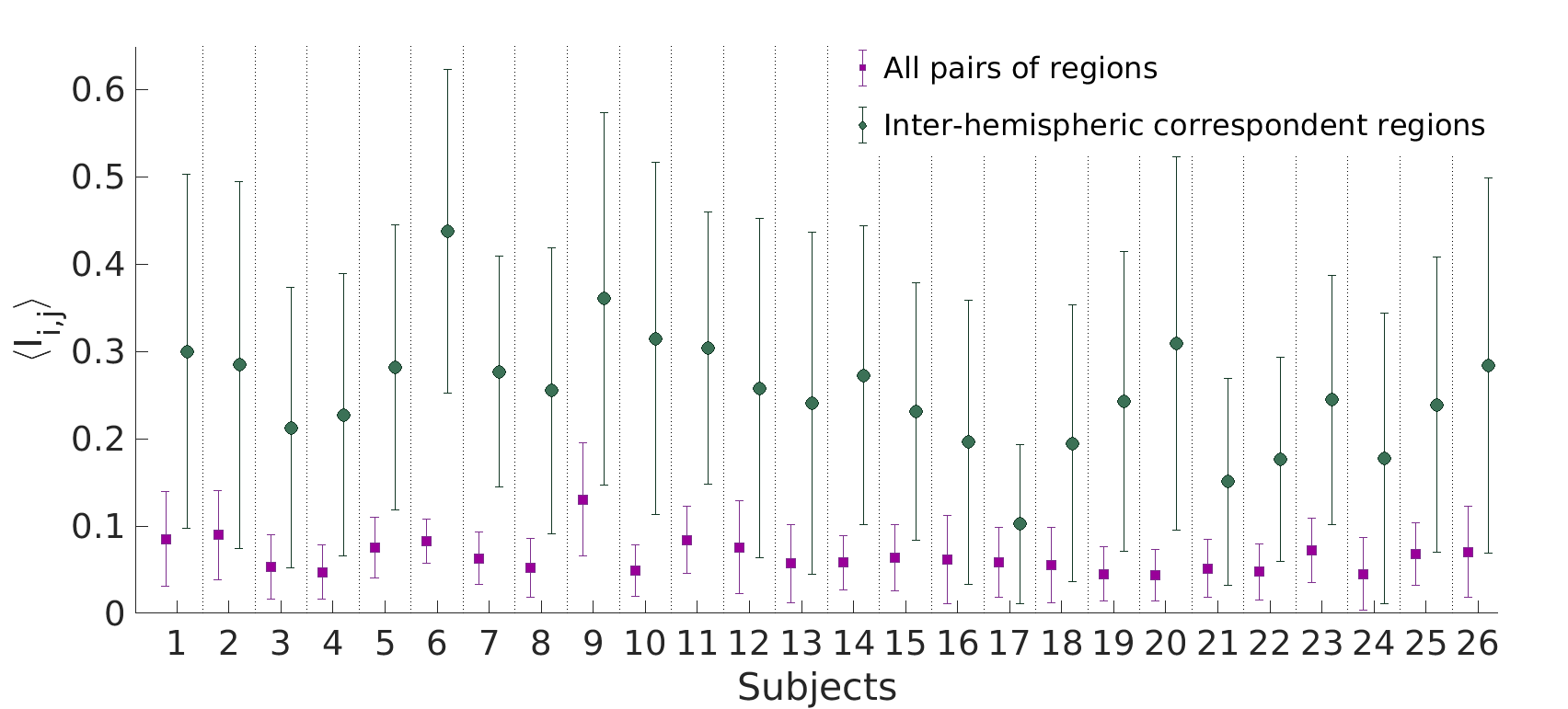}\\
	(c)
	\includegraphics[width=0.405\linewidth]{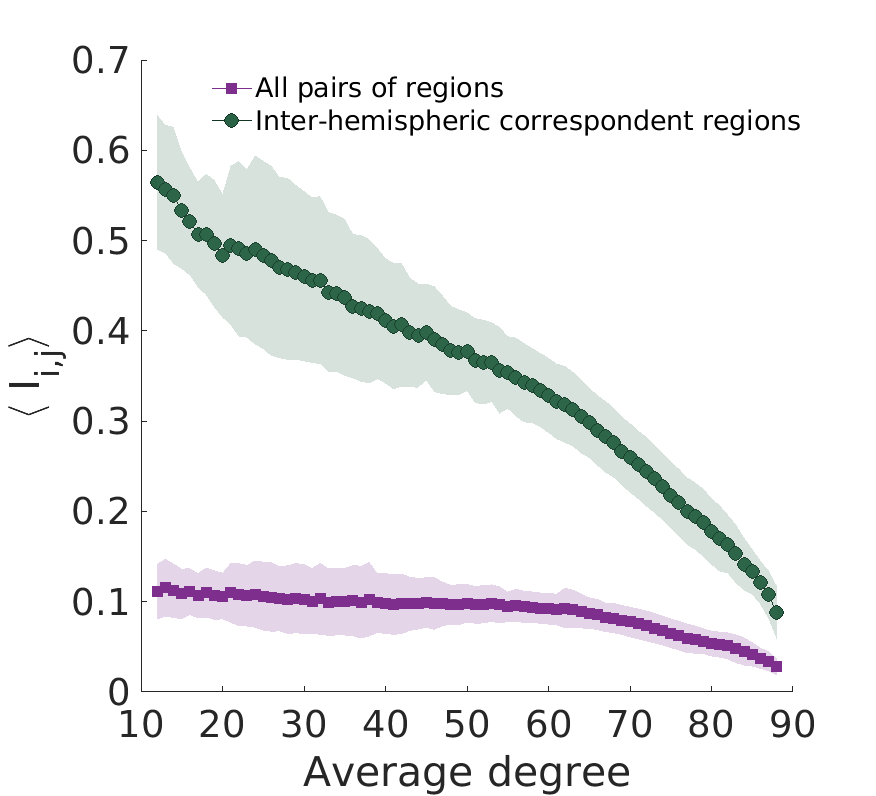}
	(d)
	\includegraphics[width=0.405\linewidth]{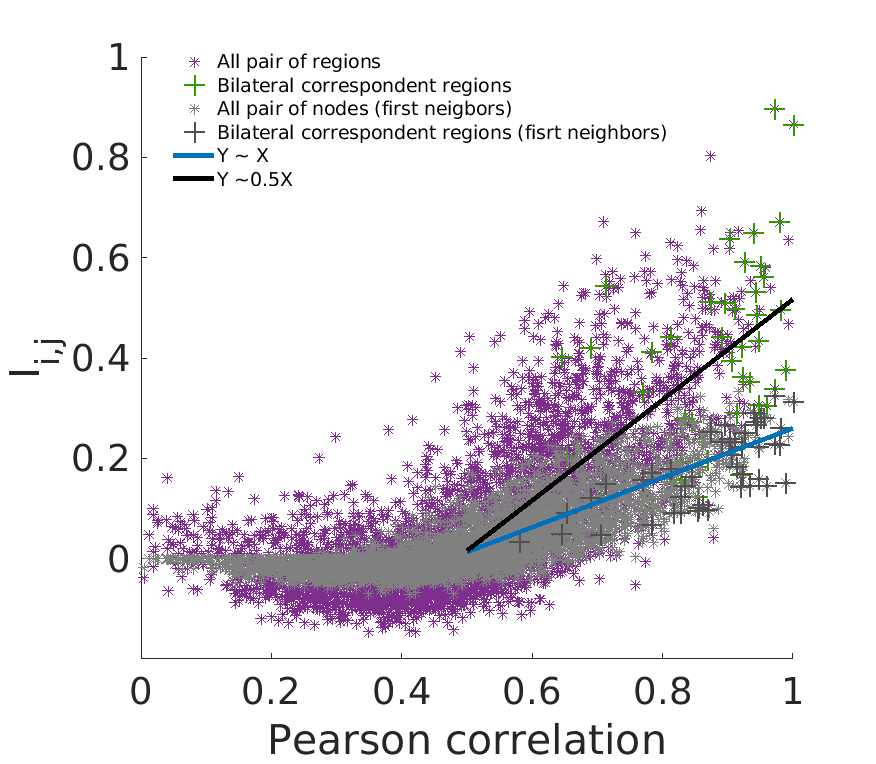}
	\begin{singlespace}\caption{
			Information parity of functional brain networks. (a) Exemplary Pearson correlation $\{C_{ij}\}$ and information parity $\{I_{ij}\}$ matrices of one subject. The homologous inter-hemispheric regions correspond to the diagonal in the lower left quadrant (black square). (b) Information parity averaged over the corresponding inter-hemispheric regions (green) and the whole brain (purple) for 26 subjects considering different average degrees, that is, different network densities. (c) Information parity as shown in panel (b), but in dependence on the average degree and averaged over all subjects. (d) Scatter of all links of one sample showing the relationship between information parity and Pearson correlation in purple; the inter-hemispheric homologous regions are highlighted in green. The information parity calculated considering only the first neighbors is shown in gray.
		}
		\label{functional}
	\end{singlespace}
	
\end{figure*} 
\section{Discussion}

Aiming to characterize nontrivial symmetries on complex networks, we have proposed a novel method based on statistics of the geodesic distances distributions.
The probability distribution of the geodesic distances reflects the diversity of influences that each node is subjected to \cite{ge}, taking into account direct and indirect influences. On the other hand, the probability distribution to find equidistant neighbors reflects the diversity of common influences for a pair of nodes. 
Information parity quantifies the congruity of influences between nodes imposed by the whole network structure, that is, how similar is the information that a pair of nodes share considering the whole network topology. 
We have selected three problems of network sciences to illustrate the potential of the information parity. 
Our analyses have indicated that individual ideological orientations in social networks are strongly influenced by the information parity between members of a community. Information parity reveals the role of each individual in the network.
%

We have also characterized bilateral symmetries evaluating information parity of homologous cortical regions in human brain networks. 
We have detected that on anatomical networks regions near the midsagittal line have a significant greater information parity. 
This can be explained by the brain anatomy or limitations of the data acquisition technique. 
We also have shown that on functional brain networks, information party is high for all inter-hemispheric homologous regions. Our results have indicated that the correlation between cortical signals is influenced by the nontrivial symmetries quantified by  information parity. 

\section{Conclusion}
Information parity is an insightful tool for the analysis of complex networks. 
The few problems explored here can be largely extended. One can explore, for example, functional correlations of the brain under task-evoked conditions or disorders and underlying brain plasticity rules. Similarly, sophisticated strategies to promote or avoid ideological polarization in social networks can be developed on the basis of the information parity. Consider, for instance, adding or removing links to a social networks based on the information parity. 
The concept of information parity has the potential to bring valuable insights for many other fields of network science as well.         

%
\clearpage
\twocolumngrid
\section*{Author contributions}
AV designed the present study and prepared all figures. VV built the functional brain networks data. All authors discussed the results and wrote the manuscript.
\vspace{1.13cm}
%

\begin{acknowledgments}
The authors acknowledge support by Deutsche Forschungsgemeinschaft under Grant No. HO4695/3-1 and within the framework of Collaborative Research Center 910. The authors thank Dr. Danielle Bassett for providing anatomical networks data. AV thank Dr. Gandhi Viswanathan and Jorge Ruiz for discussions.
\end{acknowledgments}
\bibliographystyle{apsrev4-1}
\bibliography{avio19}

\clearpage
\newpage
\onecolumngrid
\section*{Supplementary information}

\subsection{Generalized formula for information parity }
A more general formula of information parity considers different radii $r_i$ and $r_j$, which can be interpreted as independent parameters:
\begin{equation}
p_{ij}(r_i,r_j)= \frac{1}{N-2}\sum_{\substack{ k \in V \\ k\neq i,j}} \delta_{D_{ik},r_i} \, \delta_{D_{jk},r_j},
\end{equation}
%
which yields a general formula :
\begin{equation}  
I_{ij}= \sum_{r_i=1}^{r_{\mathrm{max}}} \sum_{r_j=1}^{r_{\mathrm{max}}} p_{ij}(r_i,r_j)  \log\frac{{p_{ij}(r_i,r_j)}}{p_i(r_i)p_j(r_j)}.
\end{equation}
This formula can be used to evaluate constraints and nontrivial dependencies in a network regarding to all relations. 

\subsection{Properties of the Information parity }

\begin{itemize}
	\item  The probabilities are nonzero for distance smaller than $r_{\text{max}}$ and zero otherwise. 
	
	$p(r<r_{\text{max}}) \neq 0 $\\
    $p(r\geq r_{\text{max}}) = 0 $.
	
	\item Defining the contribution $\xi_{ij}(r)=p_{ij}(r)\log\frac{p_{ij}(r)}{p_i(r)p_j(r)}$ to the mutual information $I_{ij}=\sum{ \xi_{ij}(r)}$, we defined:
	
    $p_{i}(r)=0$ or $p_{j}(r)=0 \implies \xi_{ij}(r)\equiv0$;\\
	$p_{ij}(r)=0 \implies \xi_{ij}(r)=0$;\\
    $p_{ij}(r)>p_{i}(r)p_{j}(r) \implies \xi_{ij}(r)>0$;\\
    $p_{ij}(r)<p_{i}(r)p_{j}(r) \implies \xi_{ij}(r)<0$.
   
    \item The maximum number of possibilities ($Pn$) to distribute $N-1$ nodes in $r_{\text{max}}$ neighborhood radius is given by the binomial coefficients 
\begin{equation}\nonumber
 Pn=\frac {(N-1)!}{r_{\text{max}}!(N-1-r_{\text{max}})!} ~.
\end{equation}

\end{itemize}

\newpage

\subsection{Aditional information for the Karate club network }
 \begin{figure}[ht]
	\includegraphics[width=\linewidth]{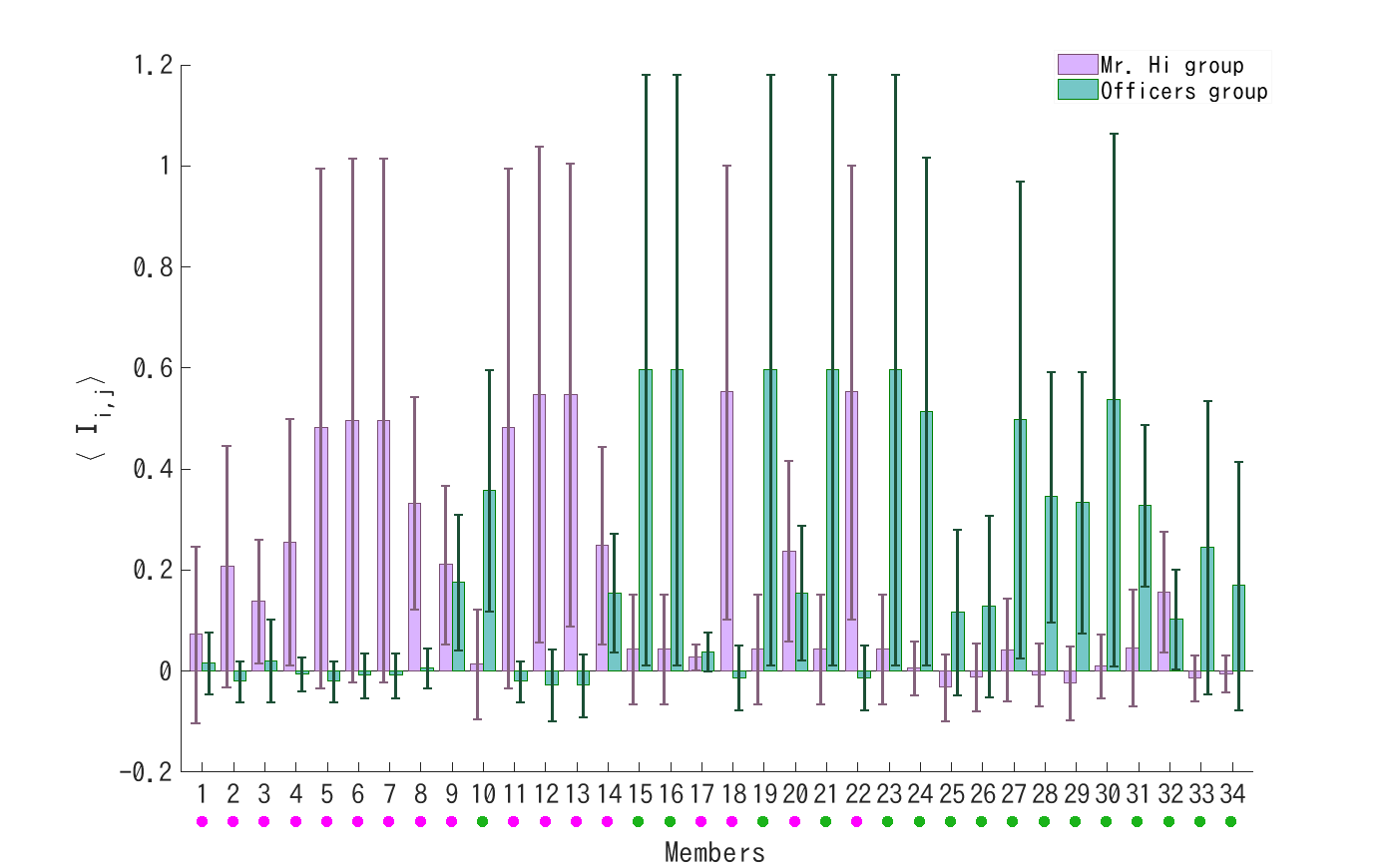}
	\caption*{Figure 1: Information parity of all Karate-club members. The pink bars depicts the average information parity of each member with Mr. Hi group and the green bars with the officers group. The error bars shows the standard deviations. The group of each member belonged after the fission are indicated by the dots following the same colors of the legend.}
\end{figure}  

\newpage

\subsection{Brain datasets}
\subsubsection*{Anatomical brain networks}
The  anatomical networks we evaluated here are the same of reference \cite{Kahn2017}; we consider only the first MRI scan section of 21 subjects  the AAL atlas for 90 regions.
The data are available on the website \url{https://complexsystemsupenn.com/}.
We normalized the weighted matrix $\{W_{ij}\}_{i,j=1,\cdots,N}$ created from diffusion imaging tractography to force it variate  from $0$ to $1$; the adjacency matrix was designed by considering $A_{ij}=1$ when  $W_{ij}$ is greater than a defined threshold. Because the qualitative results does not change according the threshold we choose the ones that yield network with average degree $\langle k \rangle = 20$.       

\subsubsection*{Functional brain networks}
The functional networks explored in this paper was obteined from fMRI data of 26 subjects in resting state with open eyes. The data are same studied in the study of the reference \cite{Vlasov2017,VUKSANOVIC2014}. The raw fMRI data are available on the 1000 Functional Connectome Project website (\url{http://fcon_1000.projects.nitrc.org/}; data ``Berlin Margulies" \cite{Biswal2010}). The functional matrices were created considering the whole frequency spectrum, without any filter.    

We create the unweighted networks thresholding the correlation matrix $\{C_{ij}\}_{i,j=1,\cdots,N}$. The adjacency matrix elements are define as $A_{ij}=1$ if the $C_{ij}$ is greater than a defined correlation threshold and  $A_{ij}=0$, otherwise. To ensure that the threshold does not influence in the results, we evaluate a range of thresholds that yield networks with different densities. The influence of the network density can be observed in the figure \ref{functional} (c). The method used here is described in the references \cite{liuyong2008,onias2014, VIO17a}.

\newpage

\subsection{List of brain regions}
	\vspace{-1.cm}
\begin{table}[h!]
	\caption{Brain regions according to the automated anatomic labelling (AAL) template. Indexes from 1-45/46-90 indicate right (R)/left (L) hemisphere.}
	\centering

	 \setlength\extrarowheight{-6pt}
	\resizebox{0.38\textwidth}{!}{
		\begin{tabular}{|l|l|r|}
			\hline
			Index R/L & Anatomical Description & Label \\
			\hline
			1/46 & Precentral & PreCG \\
			2/47 & Frontal Sup  & SFGdor \\
			3/48 & Frontal Sup Orb & ORBsup \\
			4/49 & Frontal Mid & MFG\\
			5/50 & Frontal Mid Orb & ORBmid \\
			6/51 & Frontal Inf Oper & IFGoperc\\
			7/52 & Frontal Inf Tri & IFtriang\\
			8/53 & Frontal Inf Orb & ORBinf\\
			9/54 & Rolandic Oper & ROL\\
			10/55 & Supp Motor Area & SMA \\
			11/56 & Olflactory & OLF\\
			12/57 & Frontal Sup Medial & SFGmed\\
			13/58 & Frontal Mid Orb & ORBsupmed\\
			14/59 & Gyrus Rectus & REC \\
			15/60 & Insula & INS \\
			16/61 & Cingulum Ant & ACG\\
			17/62 & Cingulum Mid & DCG\\
			18/63 & Cingulum Post & PCG\\
			19/64 & Hippocampus & HIP\\
			20/65 & ParaHippocampal & PHG\\
			21/66 & Amygdala & AMYG\\
			22/67 & Calcarine & CAL\\
			23/68 & Cuneus & CUN\\
			24/69 & Lingual & LING\\
			25/70 & Occipital Sup & SOC\\
			26/71 & Occipital Mid & MOG\\
			27/72 & Occipital Inf & IOG\\
			28/73 & Fusiform & FFG\\
			29/74 & Postcentral & PoCG\\
			30/75 & Parietal Sup & SPG\\
			31/76 & Parietal Inf & IPL\\
			32/77 & Supra Marginal Gyrus & SMG \\
			33/78 & Angular & ANG\\
			34/79 & Precuneus & PCUN\\
			35/80 & Paracentral Lobule & PCL\\
			36/81 & Caudate & CAU\\
			37/82 & Putamen & PUT\\
			38/83 & Pallidum & PAL\\
			39/84 & Thalamus & THA\\
			40/85 & Heschi & HES\\
			41/86 & Temporal Sup & STG \\
			42/87 & Temporal Pole sup & TPOsup\\
			43/88 & Temporal Mid & MTG\\
			44/89 & Temporal Pole Mid & TPOmid\\
			45/90 & Temporal Inf & ITG\\
			\hline
		\end{tabular}}
\end{table}

\end{document}